\documentstyle[12pt,aaspp4,epsfig]{article}
\begin{document}

\title{On the Frequency Evolution of X-ray Brightness Oscillations During
Thermonuclear X-ray Bursts: Evidence for Coherent Oscillations}
\author{Tod E. Strohmayer\altaffilmark{1} and 
Craig B. Markwardt\altaffilmark{1,2}}
%\affil{Laboratory for High Energy Astrophysics, Goddard Space Flight Center,
%Greenbelt, MD 20771}
\altaffiltext{1}{NASA's Goddard Space Flight Center, Code 662, Greenbelt, MD
20771; stroh@clarence.gsfc.nasa.gov, \ craigm@lheamail.gsfc.nasa.gov.}
\altaffiltext{2}{National Research Council Resident Associate.}
\authoraddr{Laboratory for High Energy Astrophysics, Mail Code 662, NASA/GSFC
Greenbelt, MD 20771}

\begin{abstract}

We investigate the time dependence of the frequency of X-ray brightness
oscillations during thermonuclear X-ray bursts from several neutron star 
low mass X-ray binaries (LMXB). We find that the oscillation frequency in
the cooling tails of X-ray bursts from 4U 1702-429 and 4U 1728-34 is well
described by an exponential ``chirp" model. With this model we demonstrate
that the pulse trains in the cooling tails of many bursts are highly phase
coherent. We measure oscillation quality factors for bursts from 4U 1728-34 
and 4U 1702-429 as high as $Q \equiv \nu_0 / \Delta\nu_{fwhm} \sim 4000$. 
We use this model of the frequency evolution to search sensitively
for significant power at the harmonics and first sub-harmonic of the 330 and 
363 Hz signal in bursts from 4U 1702-429 and 4U 1728-23, respectively,  
but find no strong evidence for significant power at any harmonic or the
subharmonic. We argue that the high coherence of the oscillations favors 
stellar rotation as the source of the oscillations. The lack of a sub-harmonic
both in bursts from 4U 1728-34 and 4U 1702-429 suggests that in these sources
the burst oscillation frequency is indeed the stellar spin frequency. 
We briefly discuss the frequency evolution in terms of rotational motion of an
angular momentum conserving thermonuclear shell. We discuss how the limits on
harmonic content can be used to infer properties of the neutron star.

\end{abstract}

\vfill\eject
%\twocolumn
%\singlespace

\section{Introduction}

Millisecond oscillations in the X-ray brightness during thermonuclear bursts,
``burst oscillations", have been observed from six low mass X-ray binaries
(LMXB) with the Rossi X-ray Timing Explorer (RXTE) 
(see \markcite{SSZ}Strohmayer, Swank \& Zhang {\it et al.} 
1998 for a review).
The presence of large amplitudes near burst onset combined with
spectral evidence for localized thermonuclear burning suggests that these
oscillations are caused by rotational modulation of thermonuclear 
inhomogeneities (see \markcite{SZS}Strohmayer, Zhang \& Swank 1997). The
asymptotic pulsation frequency in the cooling tails of bursts from 4U 1728-34
are stable over year timescales, also supporting a coherent mechanism such as
rotational modulation (\markcite{S98a}Strohmayer et al. 1998a).

An intriguing aspect of these oscillations is the frequency
evolution evident during many bursts. The frequency
is observed to increase in the cooling tail, reaching a plateau or
asymptotic limit (see \markcite{S98a}Strohmayer et al. 1998a). 
However, \markcite{S99}Strohmayer (1999) has recently discovered an episode of
spin down in the cooling tail of a burst from 4U 1636-53. 
Evidence of frequency change has been seen in
five of the six burst oscillation sources and appears to be commonly 
associated with the physical process responsible for the pulsations.
\markcite{SJGL}Strohmayer et. al (1997)
have argued this evolution results from angular momentum conservation of the
thermonuclear shell. The thermonuclear flash expands the shell,
increasing its rotational moment of inertia and slowing its spin rate.
Near burst onset the shell is thickest and thus the observed
frequency lowest. The shell then spins back up as it recouples to the
bulk of the neutron star as it cools. 
This scenario is viable as long as the shell decouples from the bulk of the
neutron star during the thermonuclear flash and then comes back into
co-rotation with it over the $\approx 10$ s of the burst fall-off. 
Calculations indicate that the $\sim 10$ m
thick pre-burst shell expands to $\sim 30$ m during the flash 
(see \markcite{Joss}Joss 1978; \markcite{B95}Bildsten 1995), 
which gives a frequency shift due to
angular momentum conservation of $\approx 2 \ \nu_{spin} (20 \ {\rm m}/ R)$,
where $\nu_{spin}$
and $R$ are the stellar spin frequency and radius, respectively. For the several
hundred Hz spin frequencies inferred from burst oscillations this gives a
shift of $\sim 2$ Hz, similar to that observed.

In bursts where frequency drift is evident the drift broadens the peak in the
power spectrum and produces quality values $Q \equiv \nu_0 / \Delta\nu_{FWHM}
\approx 300$. 
In some bursts a relatively short train of pulses is observed during 
which there is no strong evidence for a varying frequency. A burst such as
this from KS 1743-26 with 524 Hz oscillations yielded the highest coherence of 
$Q \approx 900$ yet reported in a burst oscillation (see \markcite{SMB}Smith,
Morgan \& Bradt 1997). 

In this Letter we investigate the time dependence of the
frequency observed in bursts from 4U 1728-34 and 4U 1702-429. 
We show that in the cooling tails of bursts the pulse trains are
effectively coherent. We show that with accurate modeling of the drift quality
factors as high as $Q \sim 4,000$ are achieved in some bursts. 
We investigate the functional form of the frequency drift
and show that a simple exponential ``chirp" model works remarkably well. 
We use this model to search for significant power at the harmonics and
first subharmonic of the strongest oscillation frequency in each source.
Such searches are important in establishing whether the strongest oscillation
frequency is the stellar spin frequency or its first harmonic, as appears now 
to be the case for 4U 1636-53 (see \markcite{M98}Miller 1999). The detection of
harmonic signals or limits on them is also important in obtaining constraints
on the stellar compactness (see \markcite{ML98}Miller \& Lamb 1998, 
\markcite{S98b}Strohmayer et al. 1998b). 
We note that \markcite{Z98}Zhang et al. (1998) have previously reported on a
model for the frequency evolution during a burst from Aql X-1.

\section{Modelling the Frequency Drift}

To investigate the frequency evolution in burst data we use the
$Z_{n}^2$ statistic (see \markcite{B83}Buccheri et al. 1983)
\begin{equation}
Z_n^2 = 2/N \sum_{k=1}^{n} \left ( \sum_{j=1}^N \cos (k\phi_j)\right )^2 + 
\left ( \sum_{j=1}^N \sin (k\phi_j)\right )^2 \; .
\end{equation}
Here $N$ is the total number of photons in the time series, $\phi_j$ are the
phases of each photon derived from a frequency model, $\nu (t)$, {\it vis.}
$\phi_j = 2\pi \int_0^{t_j} \nu (t') dt'$, and $n$ is the total number of
harmonics added together. For the burst oscillations,
which are highly sinusoidal, we will henceforth restrict ourselves to $n=1$. 
This statistic is particularly suited to event mode data, since no binning 
is introduced. $Z_1^2$ has the same statistical
properties as the well known Leahy normalized power spectrum, which for a
Poisson process is distributed as $\chi^2$ with 2 degrees of
freedom. All of the bursts discussed here were observed with the Proportional
Counter Array (PCA) onboard RXTE and sampled with 125 $\mu$s (1/8192 s)
resolution. 

For $\nu (t)$ we have investigated a number of functional 
forms, including; $\nu (t) = \nu_0$ (a constant frequency), $\nu (t) = 
\nu_0 (1 + d_{\nu} t)$ (a linearly increasing frequency), and
$\nu (t) = \nu_0 (1 - \delta_{\nu} \exp(-t/\tau) )$ (an exponential ``chirp"). 
For a given data set and frequency model we vary the parameters so as
to maximize the $Z_1^2$ statistic. We then compare the maximum values from
different models to judge which is superior in a
statistical sense. Our aim is to both constrain the functional form of the 
frequency evolution and to determine whether the pulse train during all
or a portion of a burst is coherent, or not. 
We judge the coherence of a given model by computing the
quality factor $Q \equiv \nu_0 / \Delta\nu_{fwhm}$ from the
width of the peak in a plot of $Z_1^2$ vs the frequency parameter $\nu_0$. 
We also compare the peak width to that expected for a coherent pulsation
in data of the same length. A pulsation in a time series of finite extent
produces a broadened peak in a power spectrum. The well known
window function, $W(\nu) = | \sin (\pi\nu T) / \pi\nu |^2$, gives a width of
$\Delta \sim 1/T$, where $T$ is the length of the data. We also confirm
that for a successful frequency model the integrated power under the $Z_1^2$
peak is consistent with that calculated assuming no frequency evolution. 

\subsection{Linear and Exponential Frequency Drift}

To begin we demonstrate how a
linear increase in frequency yields a significant improvement in the 
$Z_1^2$ statistic compared with a constant frequency model. 
We use the burst from 4U 1702-429 observed on July 26, 1997 at 14:04:19 UT, 
which we refer to as burst A (see Figure 4 in 
\markcite{MSS99}Markwardt, Strohmayer \& Swank 1999). 
We used a 5.25 s interval during this burst to investigate the frequency
evolution. In figure 1a we show results from our
calculations of $Z_1^2$ for the constant frequency model (top panel) and the
model with a linearly increasing frequency (bottom panel). In both cases the
ordinate corresponds to the frequency parameter $\nu_0$ defined in the models.
For the linear frequency model we found that $Z_1^2$ was maximized with 
$d_{\nu} = 1.264 \times 10^{-3}$ s$^{-1}$. Including the
linear drift increased $Z_1^2$ from 88.48 to 271.4, a dramatic improvement of
$\sim 183$ obtained with only 1 additional degree of freedom. The resulting
$Z_1^2$ peak is also substantially narrower (see figure 1a), leaving no doubt
that the pulsation frequency is increasing during this time interval.

The frequency evolution during bursts can also be explored with dynamic power
spectra. Several such spectra have been presented elsewhere 
(see \markcite{S98a}Strohmayer et al. 1998a,
\markcite{SSZ}Strohmayer, Swank, \& Zhang 1998, etc.). A striking behavior is
that the pulsation frequency reaches an asymptotic limit in many
bursts. Motivated by this behavior we investigated a simple
exponential ``chirp" model with a limiting frequency, 
$\nu (t) = \nu_0 (1 - \delta_{\nu} \exp(-t/\tau) )$. This model has
three parameters, the limiting frequency $\nu_0$, the fractional frequency
change, or ``bite", $\delta_{\nu}$, and the relaxation timescale, $\tau$. 
We fit this model to burst A and find a maximum $Z_1^2$ of 342.9, an increase 
of 71.5 in $Z_1^2$ over the linear frequency model. 
This is also a dramatically significant improvement in $Z_1^2$.
We fit the peak in $Z_1^2$ vs. $\nu_0$ obtained with the chirp model to
a gaussian in order to determine its width. Figure 1b shows the resulting fit.
The peak is well described by a gaussian with a width $\Delta\nu_{fwhm} = 0.201$
Hz, which gives $Q = \nu_0 / \Delta\nu_{fwhm} = 1,641$. 
We can compare this with the width caused by windowing, which for a 5.25 s
interval gives a width (FWHM) of $\approx 0.17$ Hz. 

We used the chirp model to investigate a sample of bursts
from 4U 1702-429 and 4U 1728-34. We do not present here a systematic 
description of all observed bursts, rather, we demonstrate the main results with
several illustrative examples. A burst from 4U 1702-429 observed on July 30,
1997 at 12:11:58 UT (burst B) revealed a $\sim 12$ s interval during which
oscillations were detected. Our results using the chirp model for this burst 
are summarized in figure 2. Panel (a) shows a contour plot
of the time evolution of the $Z_1^2$ statistic through the burst. It was
computed by calculating $Z_1^2$ on a grid of constant frequency values using 
2 s intervals with a new interval starting every 0.25 s, that is, assuming no
frequency evolution. The burst countrate profile (solid histogram) and best
fitting exponential chirp model (heavy solid line) are overlaid. 
The extent of the model curve defines the time interval
used to fit the chirp model. The best model tracks the dynamic $Z_1^2$
contours remarkably well. Panel (b) compares $Z_1^2$ vs. $\nu_0$ for the
constant frequency (dashed curve) and chirp models (solid curve). 
We again fit a gaussian to the peak calculated with the
chirp model and find a width $\Delta\nu_{fwhm} = 0.086$ Hz, which
yields $Q = \nu_0/\Delta\nu_{fwhm} = 3,848$ for this burst. This compares with
a width of $\approx 0.071$ for a windowed pulsation of duration 12.5 s. 

We carried out similar analyses to investigate the frequency evolution in bursts
from 4U 1728-34. We again found that the chirp model provides a remarkably 
useful description of the frequency drift. Table 1 summarizes our results using
the chirp model for several bursts from both 4U 1702-429 and 4U 1728-34. 

We find the peaks obtained with the chirp model are only
modestly broader than those expected for a coherent pulsation of the same 
length. Some of this additional width is likely due to the fact that pulsations
are not present during the entirety of each interval examined. It is also likely
that the chirp model is not the exact functional form of the frequency
evolution, this is suggested by the broader wings of the $Z_1^2$ peaks 
computed for several bursts, however, the success of such a simple model argues
strongly that the pulsations during the cooling tails of these bursts are 
phase coherent. 

\section{Harmonics and Subharmonics}

Pulsations from a rotating hotspot can be used to place constraints on neutron
star compactnesses (see \markcite{S98b}Strohmayer et al. 1998b;
\markcite{ML98}Miller \& Lamb 1998; and \markcite{M99}Miller 1999). 
The pulsation amplitude is constrained by the strength of gravitational light
deflection. An observed amplitude places an upper limit on the compactness, 
$GM/c^2 R$, because too compact stars cannot achieve the observed
modulation amplitude. Further, an upper limit on the harmonic content places a
lower limit on the compactness, since less compact stars produce more 
harmonic content, and at some limit the harmonics should become detectable. 

In some models for the kHz QPO observed in the accretion driven X-ray flux 
from neutron star LMXB, the QPO frequency separation is
closely related to the stellar spin frequency inferred from burst oscillations
(see \markcite{MLP}Miller, Lamb \& Psaltis 1998; 
\markcite{S96}Strohmayer et al. 1996). In two sources,
burst oscillations are seen with frequencies close to twice the kHz QPO
frequency separation (\markcite{W97}Wijnands et al. 1997;
\markcite{WV97}Wijnands \& van der Klis 1997; \markcite{Mvv}Mendez, 
van der Klis \& van Paradijs ). \markcite{M99}Miller (1999) has 
reported evidence for a significant 290 Hz subharmonic of the strong
580 Hz pulsation seen in 4U 1636-539 (\markcite{Z96}Zhang et al. 1996),
suggesting that the strongest signal observed during bursts may actually be the
first harmonic of the spin frequency, not the spin frequency itself. Based on
these new results and the evidence for a beat frequency interpretation it is
important to search for the subharmonic of the strongest signal detected during
bursts. 

We have shown that frequency drift during bursts can
greatly smear out the signal power. We have also shown that simple
models can recover a coherent peak. 
By modelling the drift we can make a much more sensitive search for harmonics. 
Moreover, we can coherently add signals from different bursts by first 
modelling their frequency evolution and then computing a total $Z_1^2$ by 
phase aligning each burst. We note that this procedure will also coherently 
add together power at any higher harmonics of the known signal. 
However, there will be a $\pi$ phase ambiguity of any signal at the first
subharmonic (see \markcite{M99}Miller 1999). 

We have added coherently the 330 Hz signals in all five bursts from 4U 1702-429
seen during our 1997, July observations (see \markcite{MSS}Markwardt, Strohmayer
\& Swank 1999). We fit the chirp model to oscillations in each burst and then
computed a total $Z_1^2$ by phase aligning them. Figure 3 shows the results of
this analysis. The top panel shows the total $Z_1^2$ power at 330 Hz obtained 
by adding the bursts coherently. 
The peak value is $\sim 1,400$ and
demonstrates that we have succesfully added the bursts coherently. The highest
power for any burst individually was $\sim 487$.  The two lower panels show the
power at the first and second harmonics of the 330 Hz signal. 
We find no evidence for a significant signal at these or higher harmonics. 
To search for a signal at the 165 Hz sub-harmonic we computed a total
$Z_1^2$ for each of 16 different combinations of the phases from each of the
five bursts. Since there is a 2-fold ambiguity when coherently adding a 
subharmonic signal from two bursts, with a total of 5 we have
$2^4 = 16$ possible combinations. 
We found no significant power at the subharmonic.  We performed
a similar analysis using 4 bursts from 4U 1728-34 which showed strong 
oscillations in their cooling tails, again we found no significant harmonic or
subharmonic signals. The 90 \% confidence upper limits on the signal power, 
$Z_1^2$, at the first harmonic in bursts from 4U 1702-429 and 4U 1728-34 are 
5.8 and 1.8, respectively. These correspond to lower limits on the ratio of
power at the fundamental to power at the first harmonic, $h$, of 242 and 556,
respectively.

\section{Discussion}

In this work we have concentrated on the pulsations in the cooling
tails of bursts. Bursts also show pulsations
during the rising phase (\markcite{SSZ}Strohmayer, Zhang, \& Swank 1997). 
We have not yet been able to show that the pulsations which begin near burst
onset can be phase connected to those in the cooling tail with a simple model.
To fully address this interesting question will require more sophisticated
modelling than we have employed here. We will address this question in future
work.

With the chirp model we find magnitudes of the frequency shift, 
$\nu_0\delta_{\nu}$ of $\sim 2 \ - 3$
Hz. These values are consistent with simple estimates based on angular momentum
conservation using theoretical values for the pre- and post-burst thickness of
bursting shells (\markcite{B95}Bildsten 1995). For the frequency relaxation
timescale, $\tau$, we find a range of values from 1.7 to 4 s. Interestingly,
different bursts from the same source can show markedly different decay
timescales. For example, the two bursts from 4U 1728-34 summarized in 
table 1 show similar values for $\nu_0$ and $\delta_{\nu}$, but have 
decay timescales $\tau$ which differ by almost a factor of two. 
Of these two bursts, burst C had both a substantially greater peak flux 
and fluence. This seems consistent with the idea that the frequency
increase is due to hydrostatic settling of the shell as it
radiates away its thermal energy, however, study of more bursts is
required to firmly establish such a connection. 

If the angular momentum conservation argument is correct, it implies the
existence of a shear layer in the neutron star atmosphere. In the chirp model
the total amount of phase shearing is simply, $\phi_{shear} = \nu_0\delta_{\nu}
\tau (1-e^{-T/\tau})$, where $T$ is the length of the data interval. For the 
bursts examined here we find $\phi_{shear} \sim 4 - 8 $, so that the shell 
``slips" this many revolutions over the underlying neutron star 
during the duration of pulsations. The dynamics of this shear layer
are no doubt complex. Given the physical conditions in the shell; the shear
flows are characterized by a large Reynolds number, it is likely that
dissipation of the shear velocity and recoupling will be dominated
by turbulent momentum transport. Magnetic fields may also play a role as well.
Shear layers can be unstable to Kelvin-Helmholtz
instability, however, \markcite{B98}Bildsten (1998) has suggested that the 
shear may be stabilized by either thermal buoyancy or the mean molecular 
weight contrast. We urge new theoretical investigations to explore the 
mechanisms of recoupling to determine if such a shear layer can 
survive long enough to explain the persistence of pulsations for 
$\sim 10$ s, as well as the observed relaxation timescale.

For rotational modulation of a hotspot, the ratio, $h$, of
signal power at the fundamental to that at the first harmonic is a function of
the stellar compactness (see \markcite{ML98}Miller \& Lamb 1998), so that
measurement of $h$ can be used to constrain the compactness. More compact stars
have less harmonic content in their pulses and therefore larger $h$. 
Since the pulsations in the cooling tails of bursts are likely caused by a 
broad brightness anisotropy on the neutron star surface, and not a point spot, 
it will require more realistic modelling of such an emission geometry to use 
the limits on harmonic content derived here to constrain the stellar
compactness. We will perform such modelling in future work.

\acknowledgements

\vfill\eject

\vfill\eject

\section{Figure Captions}

\noindent Figure 1a: $Z_1^2$ vs frequency parameter $\nu_0$ computed from a
5.25 s interval in burst A. The top panel was calculated with no
frequency modulation, $d_{\nu} = 0$, while the bottom panel was computed with a
linear frequency increase of magnitude $d_{\nu} = 1.264 \times 10^{-3}$
s$^{-1}$. 

\vskip 10pt

\noindent Figure 1b:  $Z_1^2$ peak and gaussian fit to burst A using the best
fitting parameters of the chirp model. The solid line shows the gaussian fit
to the peak. The derived peak centroid and width (FWHM) give $Q \equiv 
\nu_0 / \delta_{\nu} = 1,641$.

\vskip 10pt

\noindent Figure 2a: Dynamic $Z_1^2$ spectrum for burst B. The contours show
loci of constant $Z_1^2$ and were computed using 2 s intervals with a new
interval starting every 0.25 s. The calculation was done on a grid of constant
frequency points with no frequency modulation. The PCA countrate
profile is shown (solid histogram) as well as the best fitting chirp model
(heavy solid line). The interval used to fit the chirp model is
denoted by the extent of the model curve.

\vskip 10pt

\noindent Figure 2b: $Z_1^2$ vs frequency parameter $\nu_0$ computed from the
time interval during burst B marked in figure 2a. The dashed curve shows $Z_1^2$
computed with no frequency modulation, ie. $\delta_{\nu} = 0$, while the solid
curve was computed with the best fitting chirp model. 

\vskip 10 pt

\noindent Figure 3: Results of the search for harmonic signals by coherently
adding 330 Hz signals in five bursts from 4U 1702-429. The top panel shows
the total 330 Hz signal computed by adding all five bursts coherently. The
two lower panels show $Z_1^2$ in the vicinty of the 1st and 2nd harmonics of
the 330 Hz signal. There is no significant power at either harmonic.

\vfill\eject

\onecolumn

\hoffset=-0.4in
\begin{table}
\caption{Frequency evolution parameters and 90 \% confidence limits for bursts
from 4U 1728-34 and 4U 1702-429}

\begin{tabular}{cccccc}

Object & $T_{burst}$ (m/d/y at UT) & $\nu_0$ (Hz) & $\delta_{\nu} \ (10^{-3})$
& $\tau$ (s) & $Q$ \cr
\tableline
4U 1702-429 &  &  &  &  & \cr
\tableline
 A & 7/26/97 at 14:04:19 UT & $329.851 \pm 0.1$ & $7.7 \pm 0.32$ & 
$1.880 \pm 0.25$ & 1,641 \cr

 B & 7/30/97 at 12:11:58 UT & $330.546 \pm 0.02$ & $4.8 \pm 0.31$ & $4.016 \pm
0.07$ & 3,848 \cr

\tableline
4U 1728-34 &  &  &  &  & \cr
\tableline

 C & 2/16/96 at 10:00:45 UT & $364.226 \pm 0.05$ & $6.6 \pm 0.14$ 
& $3.520 \pm 0.28$ & 4,535 \cr

   & 9/22/97 at 06:42:56 UT & $364.102 \pm 0.05$ & $5.9 \pm 0.22$ 
& $1.843 \pm 0.15$ & 2,023 \cr

\tableline

\end{tabular}
\end{table}

\vfill\eject
\hoffset=0.0in

\begin{figure*}[htb] % fig1a
\centerline{\epsfig{file=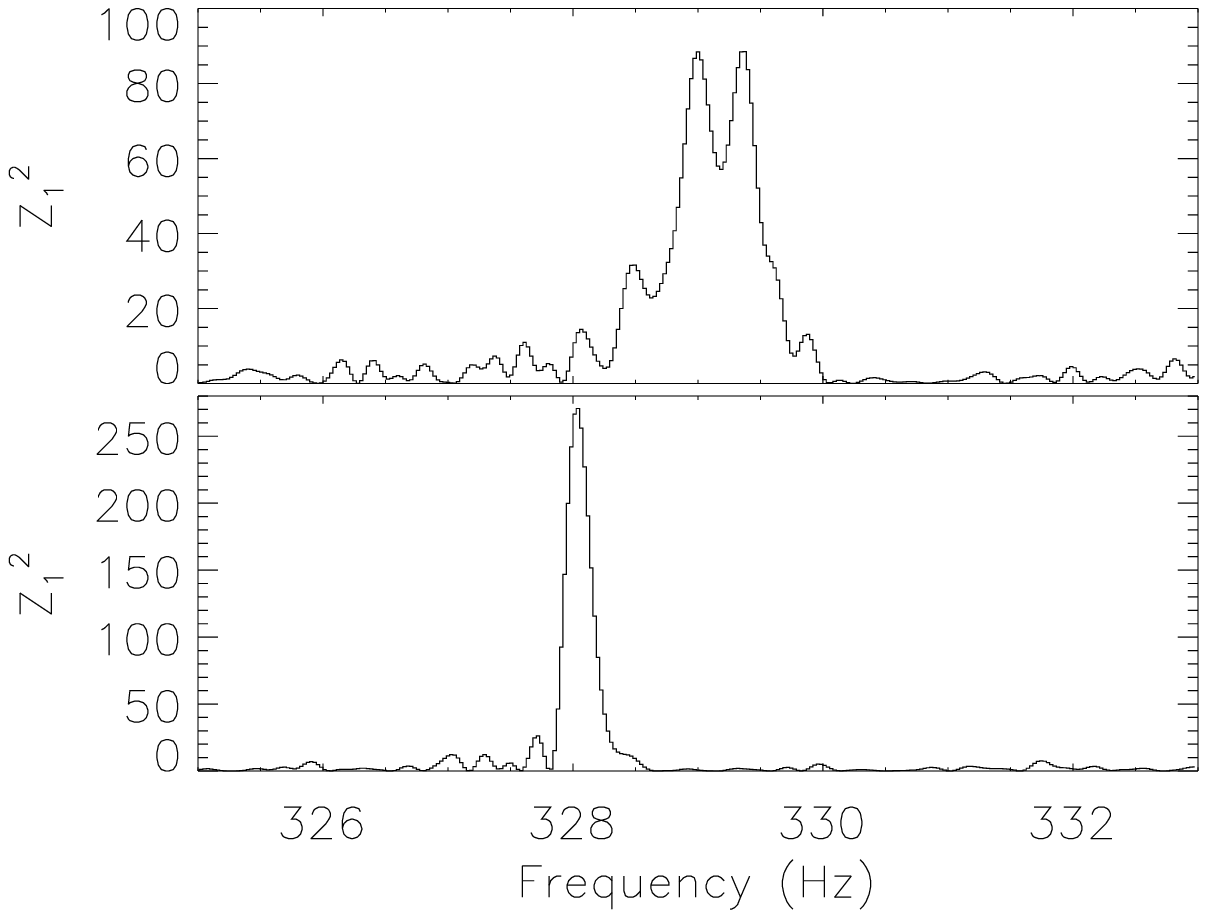,height=5.0in,width=5.0in}}
\vspace{10pt}
\caption{Figure 1a}
\end{figure*}

\vfill\eject

\begin{figure*}[htb] % fig1b
\centerline{\epsfig{file=fig2_x1702_expmodzns_gausfit.ps,height=5.0in,width=5.0in}}
\vspace{10pt}
\caption{Figure 1b}
\end{figure*}

\vfill\eject

\begin{figure*}[htb] % fig2a
\centerline{\epsfig{file=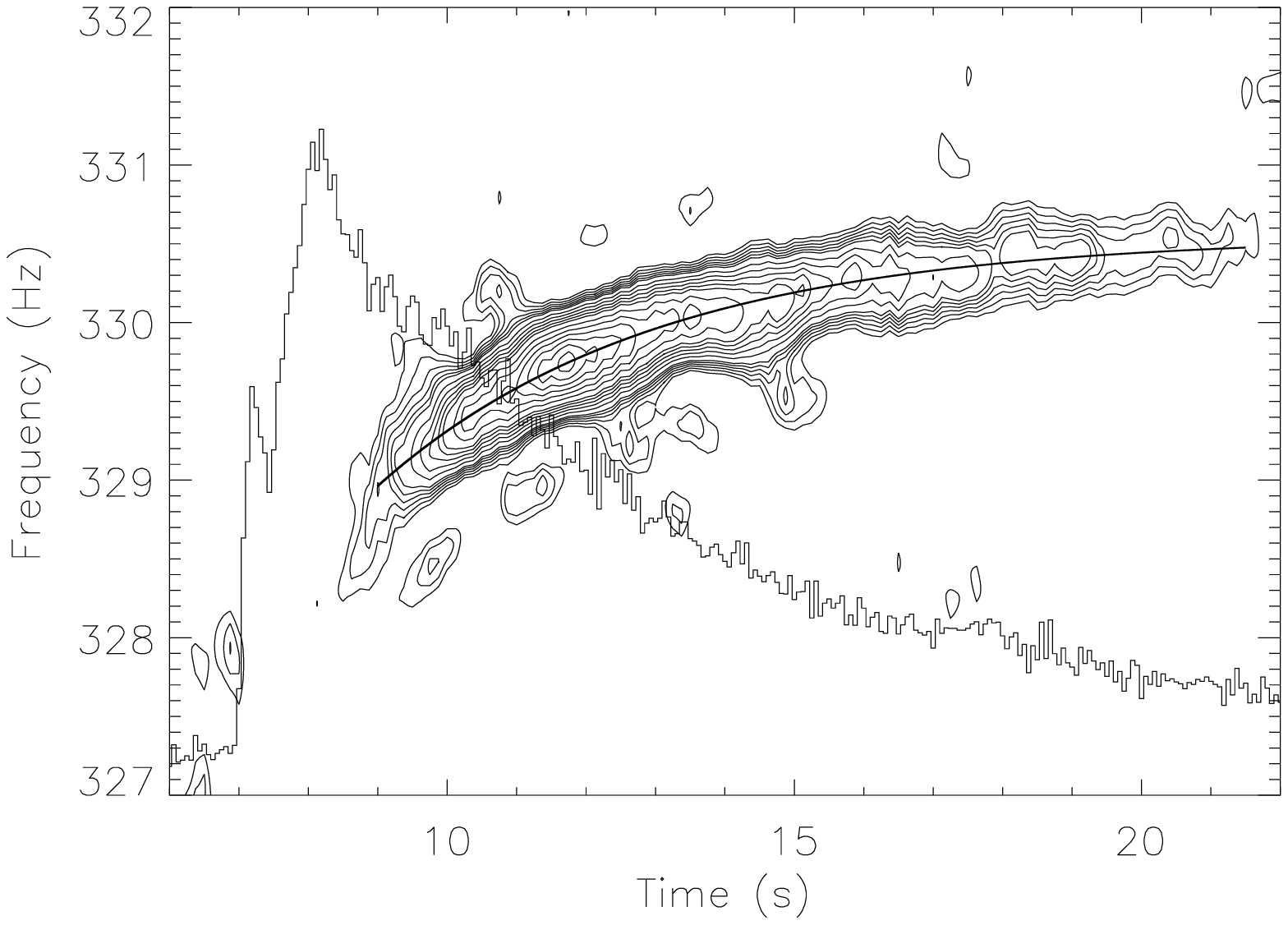,height=5.0in,width=5.0in}}
\vspace{10pt}
\caption{Figure 2a}
\end{figure*}

\vfill\eject

\begin{figure*}[htb] % fig2b
\centerline{\epsfig{file=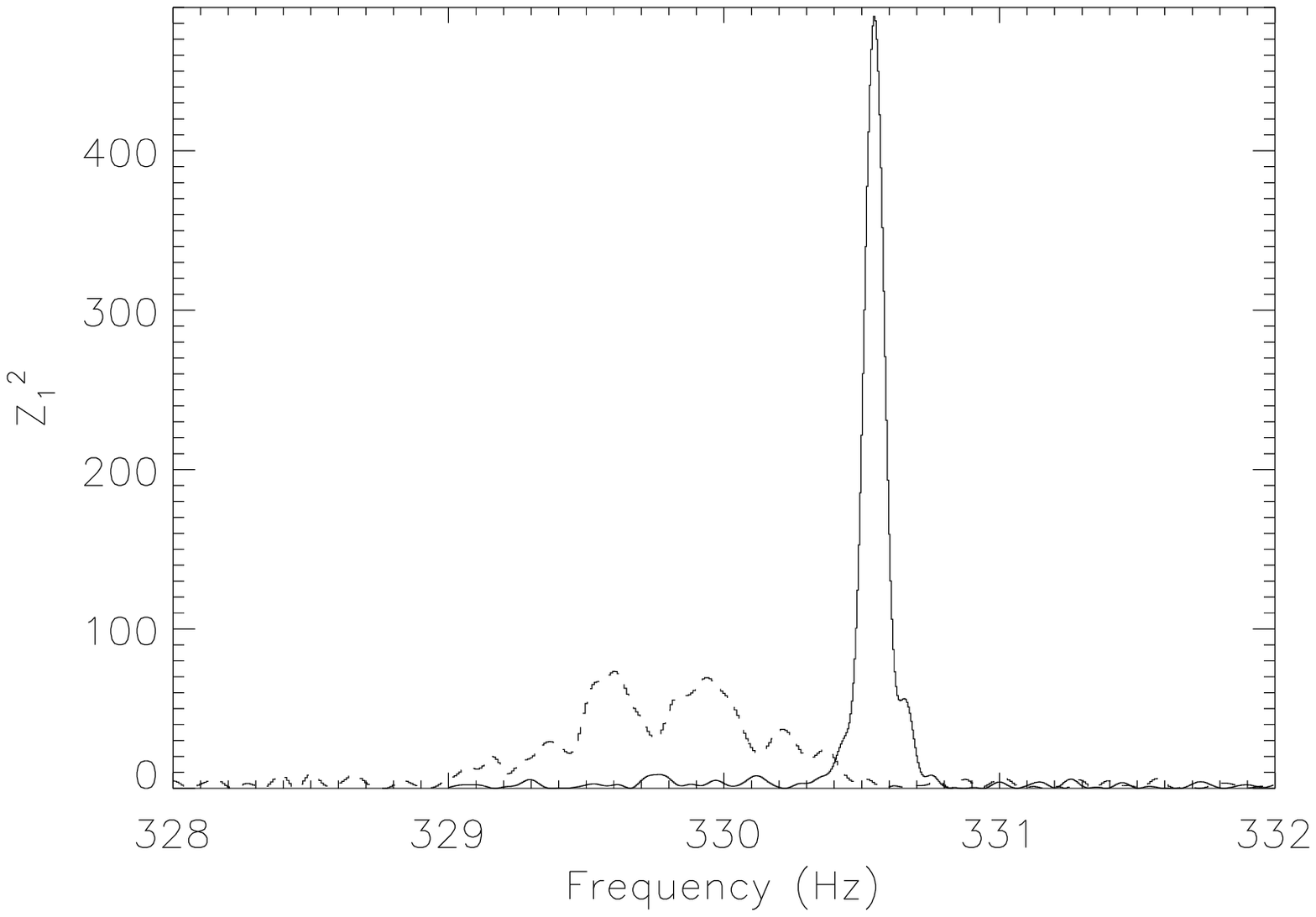,height=5.0in,width=5.0in}}
\vspace{10pt}
\caption{Figure 2b}
\end{figure*}

\vfill\eject

\begin{figure*}[htb] % fig3
\centerline{\epsfig{file=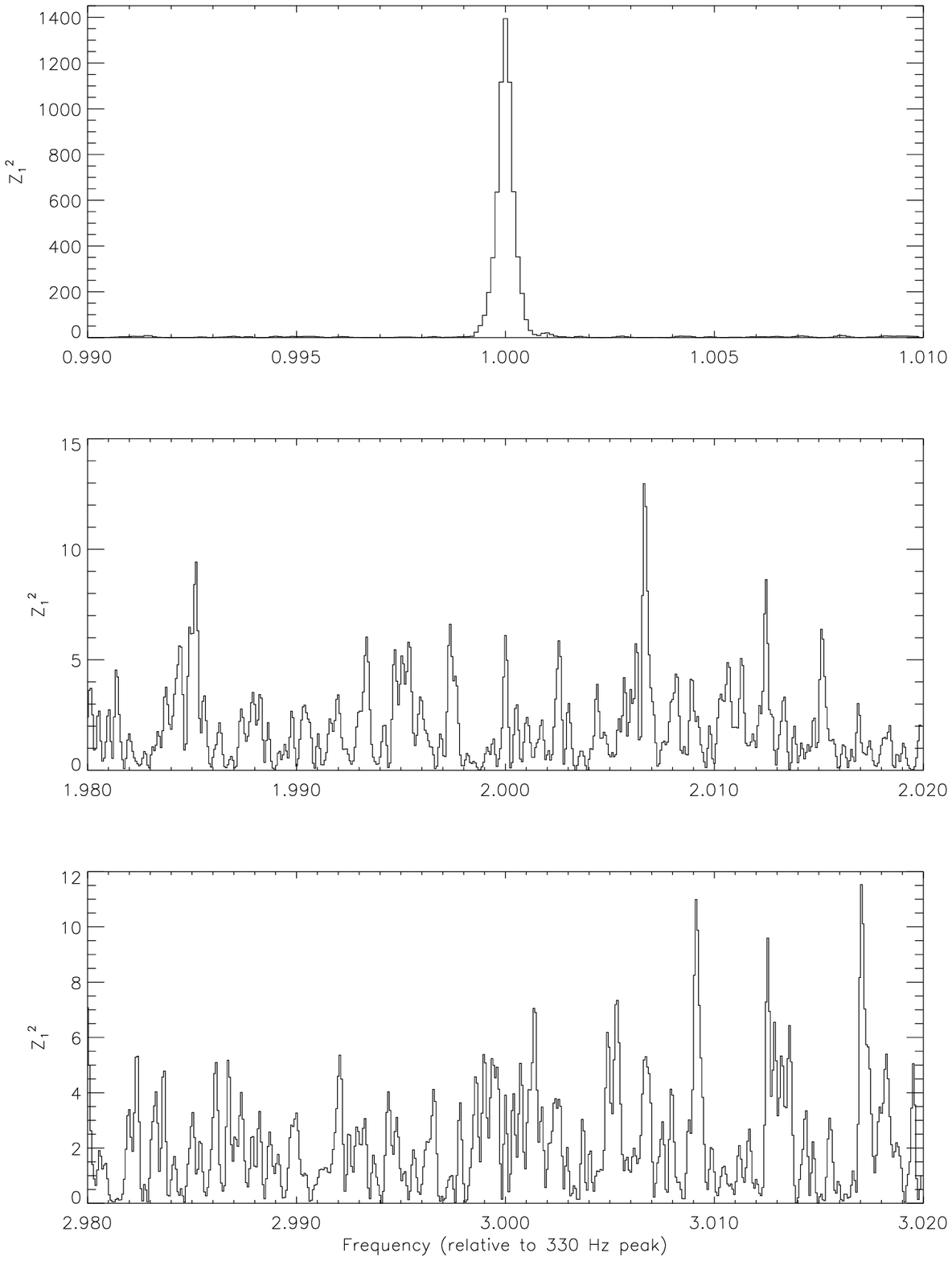,height=7.0in,width=5.0in}}
\vspace{10pt}
\caption{Figure 3}
\end{figure*}

\vfill\eject

\end{document}